\documentclass[11pt,fleqn,twoside]{article}
\usepackage{amsfonts,amssymb,latexsym}
\makeatletter
\newcommand{\prava}[1]{\small\it
\begin{flushleft}
Copyright \copyright \ 1999 by  #1
\end{flushleft}}

\newcommand{\name}[1]{\begin{flushleft}
                       \LARGE \bf #1
                       \end{flushleft}\vspace{-3mm}}

\newcommand{\Author}[1]{\begin{flushleft}
                       \it #1 \end{flushleft}}

\newcommand{\Adress}[1]{\begin{flushleft}
                       \it #1 \end{flushleft}}

\newcommand{\Date}[1]{\begin{flushleft}
                      \small  \it #1 \end{flushleft}}

\newcommand{\ehkol}{Author \ name}
\newcommand{\ohkol}{Article \ name}
\renewcommand{\@evenhead}{
\hspace*{-3pt}\raisebox{-15pt}[\headheight][0pt]{\vbox{\hbox to \textwidth 
{\thepage \hfil \ehkol}\vskip4pt \hrule}}}
\renewcommand{\@oddhead}{
\hspace*{-3pt}\raisebox{-15pt}[\headheight][0pt]{\vbox{\hbox to \textwidth 
{\ohkol \hfil \thepage}\vskip4pt\hrule}}}
\renewcommand{\@evenfoot}{}
\renewcommand{\@oddfoot}{}

\newcommand{\be}{\begin{equation}}
\newcommand{\ee}{\end{equation}}
\newcommand{\ba}{\hspace*{-5pt}\begin{array}}
\newcommand{\ea}{\end{array}}

\makeatother

\begin{document}

\thispagestyle{empty}
\setcounter{page}{255}
\renewcommand{\ehkol}{S.Yu. Sakovich}
\renewcommand{\ohkol}{Coupled KdV Equations of Hirota-Satsuma Type}

\begin{flushleft}
\footnotesize \sf
Journal of Nonlinear Mathematical Physics \qquad 1999, V.6, N~3,
\pageref{sakovich-fp}--\pageref{sakovich-lp}.
\hfill {\sc Letter}
\end{flushleft}

\vspace{-5mm}

\renewcommand{\footnoterule}{} 
{\renewcommand{\thefootnote}{} \footnote{\prava{S.Yu. Sakovich}}}

\name{Coupled KdV Equations of Hirota-Satsuma Type}\label{sakovich-fp}

\Author{S.Yu. SAKOVICH}

\Adress{Institute of Physics, National Academy of Sciences, P.O. 72,
Minsk, Belarus\\[1mm]
E-mail: sakovich@dragon.bas-net.by}

\Date{Received January 19, 1999; Revised April 24, 1999; Accepted
May 26, 1999}

\begin{abstract}
\noindent
It is shown that the system of two coupled Korteweg-de Vries equations passes
the Painlev\'{e} test for integrability in nine distinct cases of its
coef\/f\/icients. The integrability of eight cases is verif\/ied by direct
construction of Lax pairs, whereas for one case it remains unknown.
\end{abstract}

\section{Introduction}

Recently, Karasu \cite{Kar} proposed a Painlev\'{e} classif\/ication of coupled
KdV equations. More recently, we found that the classif\/ication of Karasu
missed at least two systems which possessed the Painlev\'{e} property and Lax
pairs~\cite{Sak}. In the present paper, we give our ver\-sion of the singularity
analysis of coupled KdV equations.

We study the following class of nonlinear systems of partial dif\/ferential
equations:
\be
\ba{l}
u_{xxx}+auu_{x}+bvu_{x}+cuv_{x}+dvv_{x}+mu_{t}+nv_{t}    =0,
\vspace{1mm}\\
v_{xxx}+euu_{x}+fvu_{x}+guv_{x}+hvv_{x}+pu_{t}+qv_{t}   =0,
\ea \label{1}
\ee
where $a$, $b$, $c$, $d$, $e$, $f$, $g$, $h$, $m$, $n$, $p$ and $q$ are
arbitrary constants, and
\begin{equation}
mq\neq np. \label{2}
\end{equation}
The condition (\ref{2}) allows every system (\ref{1}) to be resolved with
respect to $u_{t}$ and $v_{t}$, therefore our class (\ref{1}) coincides with
the class of all ``nondegenerate KdV systems'' studied by Karasu \cite{Kar}.
Since the ``degenerate KdV systems'' \cite{Kar} are in fact some
over-determined systems \cite{Olv} reducible to lower-order ones, we do not
classify them as coupled Korteweg-de Vries equations.

The system (\ref{1}) is ``soft'' in the sense that its coef\/f\/icients can be
changed by the transformation
\begin{equation}
u^{\prime}=y_{1}u+y_{2}v, \qquad v^{\prime}=y_{3}u+y_{4}v,
\qquad t^{\prime}=y_{5}t, \label{3}
\end{equation}
where $y_{1}$, $y_{2}$, $y_{3}$, $y_{4}$ and $y_{5}$\ are constants,
$y_{1}y_{4}\neq y_{2}y_{3}$, and $y_{5}\neq0$. At every stage of our analysis,
we use the transformation (\ref{3}) in order to simplify the system (\ref{1})
as far as possible.

We perform the singularity analysis of the class (\ref{1}) in accordance with
the Weiss-Kruskal algorithm, which is based on the Weiss-Tabor-Carnevale
singular expansions~\cite{WTC}, Ward's requirement not to analyze solutions at
their characteristics \cite{War}, Kruskal's simplifying representation of
singularity manifolds \cite{JKM}, and which follows step by step the
Ablowitz-Ramani-Segur algorithm for ordinary dif\/ferential equations
\cite{ARS}. Since the Weiss-Kruskal algorithm is well known and widely used,
we omit all unessential details of our computations.

It should be stressed, however, that the Weiss-Kruskal algorithm checks only
necessary conditions for an equation to possess the Painlev\'{e} property.
Moreover, as explained in Sections \ref{branches} and \ref{incompleteness},
our analysis could miss some systems with the Painlev\'{e} property. For these
reasons, our work does not claim to be a ``Painlev\'{e} classif\/ication'', and
we prefer to think that it is simply a search for those systems (\ref{1})
whose analytic properties are similar to such of the Hirota-Satsuma system
\cite{HS} studied in \cite{RDG}.

In Section \ref{analysis}, we study the analytic properties of the systems
(\ref{1}) and f\/ind, up to arbitrary transformations (\ref{3}), the following
nine coupled KdV equations of Hirota-Satsuma type:
\renewcommand{\theequation}{{\rm i}}
\begin{equation}
u_{t}=u_{xxx}-12uu_{x}+24kvv_{x}, \qquad v_{t}=-2v_{xxx}+12uv_{x},
\qquad k=-1,0,1, 
\label{i}
\end{equation}
\renewcommand{\theequation}{{\rm ii}}
\begin{equation}
u_{t}=u_{xxx}-12uu_{x}, \qquad v_{t}=v_{xxx}-6uv_{x},
\label{ii}
\end{equation}
\renewcommand{\theequation}{{\rm iii}}
\begin{equation}
u_{t}=u_{xxx}-12uu_{x}, \qquad v_{t}=4v_{xxx}-6vu_{x}-12uv_{x},
 \label{iii}
\end{equation}
\renewcommand{\theequation}{{\rm iv}}
\be
\ba{l}
u_{t}    =u_{xxx}-9v_{xxx}-12uu_{x}+18vu_{x}+18uv_{x},
\vspace{1mm}\\
v_{t}    =u_{xxx}-5v_{xxx}-12uu_{x}+12vu_{x}+6uv_{x}+18vv_{x},
\label{iv}
\ea
\ee
\renewcommand{\theequation}{{\rm v}}
\be
\ba{l}
u_{t}    =u_{xxx}+9v_{xxx}-12uu_{x}-18vu_{x}-18uv_{x}+108vv_{x},
\vspace{1mm}\\
v_{t}    =u_{xxx}-11v_{xxx}-12uu_{x}+12vu_{x}+42uv_{x}+18vv_{x},
\ea \label{v}
\ee
\renewcommand{\theequation}{{\rm vi}}
\begin{equation}
u_{t}=u_{xxx}-12uu_{x}, \qquad v_{t}=v_{xxx}-12uv_{x},
\label{vi}
\end{equation}
\renewcommand{\theequation}{{\rm vii}}
\begin{equation}
u_{t}=u_{xxx}-12uu_{x}, \qquad
v_{t}=v_{xxx}-6vu_{x}-6uv_{x},
\label{vii}
\end{equation}
\renewcommand{\theequation}{{\rm viii}}
\begin{equation}
u_{t}=u_{xxx}-12uu_{x}, \qquad v_{t}=-2v_{xxx}+12vu_{x}+12uv_{x},
\label{viii}
\end{equation}
\renewcommand{\theequation}{{\rm ix}}
\begin{equation}
u_{t}=u_{xxx}-12uu_{x}, \qquad v_{t}=ku_{xxx}+v_{xxx}-12vu_{x}
-12uv_{x}, \qquad k=0,1.
\label{ix}
\end{equation}

In Section \ref{integrability}, we f\/ind transformations between some of the
systems (\ref{i})--(\ref{ix}), and then prove the integrability of
(\ref{i})--(\ref{ix}) except (\ref{v}) by direct construction of Lax pairs.

In Section \ref{conclusion}, we compare our results with the results of Karasu
\cite{Kar} and give some unsolved problems.

\section{Singularity analysis\label{analysis}}

\subsection{Generic branches\label{generic}}

\subsubsection{Leading exponents\label{branches}}

Starting the Weiss-Kruskal algorithm, we determine that a hypersurface
$\phi(x,t)=0$ is non-characteristic \cite{Olv} for the system (\ref{1}) if
$\phi_{x}\neq0$; without loss of generality we take $\phi_{x}=1$, which
simplif\/ies computations and excludes characteristic singularity manifolds from
our consideration. Since (\ref{1}) is a normal system \cite{Olv}, its general
solution must contain six arbitrary functions of one variable. The
substitution of
\renewcommand{\theequation}{\arabic{equation}}
\setcounter{equation}{3}
\begin{equation}
u=u_{0}(t)\phi^{\alpha}+\cdots +u_{r}(t)\phi^{r+\alpha}+
\cdots, 
\qquad v=v_{0}(t)\phi^{\beta}+\cdots +v_{r}(t)\phi^{r+\beta}+
\cdots \label{4}
\end{equation}
with $u_{0}v_{0}\neq0$ into the system (\ref{1}) determines branches
(i.e. the sets of admissible $\alpha$, $\beta$, $u_{0}$ and $v_{0}$)
and positions $r$ of resonances for those branches. We require that
the system (\ref{1}) admits at least one singular generic branch,
where $\alpha<0$ or $\beta<0$, the number of resonances is six, and
$r\geq0$ for f\/ive of them. This requirement seems to be more
restrictive than the Painlev\'{e} property itself, therefore some
coupled KdV equations possessing the Painlev\'{e} property may be
missed.

Throughout Section \ref{generic}, we analyze only the singular generic
branches; all other  bran\-ches are considered in Section \ref{nongeneric}. Since
the number of resonances is to be six, the set of leading terms of (\ref{1})
must include $u_{xxx}$ and $v_{xxx}$, and the admissible values of $\alpha$
and $\beta$ are as follow, depending on which of the other terms of (\ref{1})
are leading and which of the coef\/f\/icients are~zero:

\vspace{-3mm}

\begin{itemize}
\item $\alpha=-2$ and $\beta$ is arbitrary;

\vspace{-2mm}

\item $\alpha=-4$ and $\beta=-6$;

\vspace{-2mm}

\item  and the same with $\alpha\rightleftarrows\beta$, which can be omitted
due to (\ref{3}).
\vspace{-3mm}

\end{itemize}

We reject systems (\ref{1}) with $\alpha=-4$ and $\beta=-6$ because of bad
positions of resonances, $r=-1, 6, 8, 12$ and $\frac{1}{2}(11\pm
i\sqrt{159})$, and proceed to systems (\ref{1}) with $\alpha=-2$ and any
integer~$\beta$.

\subsubsection{Systems with {\mathversion{bold}$\beta>-2$}\label{done}}

If $\alpha=-2$ and $\beta>-2$, then $e=0$ in (\ref{1}); moreover, $p=0$ if
$\beta>0$. Using (\ref{3}), we make $a=-12$ in (\ref{1}), and then f\/ind that
$u_{0}=1$ and
\begin{equation}
(\beta(\beta-1)(\beta-2)-2f+\beta g)v_{0}=2\delta_{0\beta}p\phi_{t}.
\label{5}
\end{equation}
Positions of resonances are $-1$, $4$, $6$, $r_{1}$, $r_{2}$ and $r_{3}$,
which correspond to the possible arbitrariness of $\phi$, $u_{4}$, $u_{6}$,
$v_{r_{1}}$, $v_{r_{2}}$ and $v_{r_{3}}$, where $r_{1}$, $r_{2}$ and $r_{3}$
are three roots of
\begin{equation}
r^{3}+3(\beta-1)r^{2}+(3\beta^{2}-6\beta+2+g)r+2\delta_{0\beta}p\phi_{t}
v_{0}^{-1}=0. \label{6}
\end{equation}
It follows from (\ref{6}) that $r_{1}+r_{2}+r_{3}=3(1-\beta)$. Since $r_{1}$,
$r_{2}$ and $r_{3}$ are to be distinct non-negative integers, it is necessary
that $\beta\leq0$. Thus, we have $\{r_{1}, r_{2}, r_{3}\}=\{0, 1, 2\}$
and $p=0$ for $\beta=0$, $\{r_{1}, r_{2}, r_{3}\}=\{0, 1, 5\}$ or
$\{0, 2, 4\}$ for $\beta=-1$, and $f=\frac{1}{2}\beta((\beta
-1)(\beta-2)+g)$ from (\ref{5}).

If $\beta=0$, then $r=-1, 0, 1, 2, 4$ and $6$, $f=0$, and $g=0$ due to
(\ref{6}). We make $b=0$ by~(\ref{3}), substitute the expansions $u=\sum\limits
_{i=0}^{\infty}u_{i}(t)\phi^{i-2}$ and $v=\sum\limits_{i=0}^{\infty}v_{i}(t)\phi^{i}$
into (\ref{1}), f\/ind recursion relations for $u_{i}$ and $v_{i}$, and then
check compatibility conditions arising at resonances. We get $c=0$ at $r=4$
and $d=n=0$ at $r=6$, therefore this case of (\ref{1}) is equivalent to two
non-coupled equations.

When $\beta=-1$ and $r=-1, 0, 1, 4, 5$ and $6$, we have $f=0$ and
$g=-6$. Then (\ref{3}) lets us make $c=-2b$ to simplify computations. We
substitute $u=\sum\limits_{i=0}^{\infty}u_{i}(t)\phi^{i-2}$ and $v=\sum\limits_{i=0}
^{\infty}v_{i}(t)\phi^{i-1}$ into (\ref{1}), and compatibility conditions at
resonances give us $h=p=b=n=0$, $d(m+2q)=0$ and $(m+2q)(m-q)=0$. Eliminating
free parameters by (\ref{3}), we get (\ref{i}) and (\ref{ii}).

When $\beta=-1$ and $r=-1, 0, 2, 4, 4$ and $6$, we have $f=-\frac
{3}{2}$ and $g=-3$, and make $c=-2b$ by (\ref{3}) again. Then compatibility
conditions at resonances give us $d=\frac{2}{3}h(4h-b)$, $q=\frac{1}
{6}p(b-4h)+\frac{1}{4}m$, $n=2bq$, $hz=pz=0$ and $h^{2}=b^{2}$, where
$z=m+\frac{2}{3}bp-\frac{16}{9}hp$. Here we have three distinct cases, which,
after simplif\/ication by (\ref{3}), become (\ref{iii}), (\ref{iv}) and (\ref{v}).

\subsubsection{Systems with {\mathversion{bold}$\beta=-2$}}

When $\alpha=\beta=-2$, $u_{0}$ and $v_{0}$ are determined by the system
\be
\ba{l}
12u_{0}+au_{0}^{2}+(b+c)u_{0}v_{0}+dv_{0}^{2}    =0,
\vspace{1mm}\\
12v_{0}+eu_{0}^{2}+(f+g)u_{0}v_{0}+hv_{0}^{2}    =0.
\ea \label{7}
\ee
If both $u_{0}$ and $v_{0}$ are some f\/ixed constants (no resonances with
$r=0$), then (\ref{3}) lets us make $\beta>-2$, but we have already considered
those cases in Section \ref{done}. The assumption that both $u_{0}$ and
$v_{0}$ are arbitrary and independent (two resonances with $r=0$) leads to a
contradiction with (\ref{7}). Therefore one of the variables $u_{0}$ and
$v_{0}$ should be arbitrary (one resonance with $r=0$), and another one should
be a function of it. According to (\ref{7}), this is possible only if $d=e=0$.
Then, using (\ref{3}), we make $a=-12$ and $c=-b$, and~(\ref{7}) gives us
$h=0$, $g=-12-f$, $u_{0}=1$, $\forall v_{0}(t)$. Positions of resonances turn
out to be $r=-1, 0, 4, 6$, $r_{1}$ and $r_{2}$, where $r_{1}$ and
$r_{2}$ are two roots of
\begin{equation}
r^{2}-9r+14-f+bv_{0}(t)=0.
 \label{8}
\end{equation}
Positions of resonances should not depend on $v_{0}(t)$, therefore we take
$b=0$, and (\ref{8}) gives us $r_{2}=9-r_{1}$ and $f=r_{1}^{2}-9r_{1}+14$,
where $r_{1}=1, 2, 3$ or $4$ because $0<r_{1}<r_{2}$.

When $r_{1}=1$, compatibility conditions at resonances lead to $m=n=q=0$,
which violates (\ref{2}), and (\ref{1}) is not a system of evolution equations
in this case.

When $r_{1}=2$, compatibility conditions at resonances give us $n=p=0$ and
$q=m$. Simplifying (\ref{1}) by (\ref{3}), we get (\ref{vi}).

When $r_{1}=3$, we f\/ind at resonances that $n=q=0$, which is prohibited by
(\ref{2}).

When $r_{1}=4$, compatibility conditions give us $n=p=q-m=0$ or $n=q+\frac
{1}{2}m=0$, and this leads through (\ref{3}) to (\ref{vii}) and (\ref{viii}).

\subsubsection{Systems with {\mathversion{bold}$\beta<-2$}\label{incompleteness}}

When $\alpha=-2$ and $\beta=-3$, we have $b=c=d=h=0$ in (\ref{1}), make
$a=-12$ by (\ref{3}), and get $u_{0}=1$, $f=-30-\frac{3}{2}g$, $\forall
v_{0}(t)$. Positions of resonances are $r=-1,0, 4, 6$, $r_{1}$ and
$r_{2}$, where $r_{1}+r_{2}=12$ and $r_{1}r_{2}=47+g$. Since $r_{1}$ and
$r_{2}$ correspond to the possible arbitrariness of $v_{r_{1}}$ and $v_{r_{2}%
}$, we have to study the following f\/ive distinct cases: $r_{1}=1, 2, 3,4$ and $5$. 
In the cases $r_{1}=1, 2$ and $3$, compatibility conditions at
resonances violate~(\ref{2}). When $r_{1}=4$, compatibility conditions
restrict arbitrary functions in expansions (\ref{4}). But in the case
$r_{1}=5$, when $f=g=-12$, we have only $n=0$ and $q=m$ at resonances, and
then, simplifying (\ref{1}) by (\ref{3}), get (\ref{ix}).

When $\alpha=-2$ and $\beta=-4$, we have $b=c=d=h=0$ in (\ref{1}). In this
case, $nv_{t}$ is a leading term, therefore we have to consider $n=0$ and
$n\neq0$ separately. There are seven distinct cases of positions of resonances
when $n=0$ and six such cases when $n\neq0$. In every case of those thirteen
cases, however, compatibility conditions at resonances either contradict
(\ref{2}) or restrict arbitrary functions in (\ref{4}).

When $\alpha=-2$ and $\beta\leq-5$, we have $b=c=d=h=n=0$ in (\ref{1}). There
are $\frac{1}{2}-\frac{3}{2}\beta$ distinct cases of positions of resonances
if $\beta$ is odd and $1-\frac{3}{2}\beta$ such cases if $\beta$ is even. Now,
using the \textit{Mathematica} computer system \cite{Wol}, we can check that,
for $\beta=-5, -6, \ldots , -10$ and for all possible positions of
resonances, compatibility conditions at resonances either contradict (\ref{2})
or restrict arbitrary functions in (\ref{4}). This is very suggestive that in
fact no systems (\ref{1}) with $\beta<-3$ possess the Painlev\'{e} property.
But we were unable to prove this conjecture within the Weiss-Kruskal algorithm
for all $\beta<-3$, and therefore our list (\ref{i})--(\ref{ix}) may be incomplete.

\subsection{Non-generic branches\label{nongeneric}}

The systems (\ref{i})--(\ref{ix}) admit many branches. The singular generic
branches have been studied in Section \ref{generic}. All the nonsingular
branches correspond to Taylor expansions governed by the Cauchy-Kovalevskaya
theorem because the system (\ref{1}) is written in a nonsingular Kovalevskaya
form \cite{Olv}. Most of admissible singular non-generic branches of
(\ref{i})--(\ref{ix}) correspond in fact to the singular generic expansions,
where one or two arbitrary functions at resonances are taken to be zero.
Therefore we have to study only the following singular non-generic branches:

\vspace{-2mm}

\begin{itemize}
\item $\alpha=\beta=-2$, $u_{0}=2$, $v_{0}^{2}=1/k$, $r=-2, -1, 3, 4,6$ and 
$8$ for (\ref{i}) with $k\neq0$;

\vspace{-2mm} 

\item $\alpha=\beta=-2$, $u_{0}=2$, $v_{0}=-\frac{2}{3}$, $r=-2, -1, 3,4, 6$ 
and $8$ for (\ref{iv});

\vspace{-2mm} 

\item $\alpha=\beta=-2$, $u_{0}=6$, $v_{0}=\frac{10}{3}$, $r=-5, -1, 4,6, 6$ and $8$ for (\ref{iv});

\vspace{-2mm} 

\item $\alpha=\beta=-2$, $u_{0}=2$, $v_{0}=-\frac{2}{3}$, $r=-2, -1, 4,
5, 6$ and $6$ for (\ref{v});

\vspace{-2mm} 

\item $\alpha=\beta=-2$, $u_{0}=4$, $v_{0}=\frac{4}{3}$, $r=-4, -1, 3,4, 6$ and $10$ for (\ref{v}).
\vspace{-2mm}

\end{itemize}

Compatibility conditions at all resonances of these branches turn out to be
satisf\/ied iden\-tically. Consequently, the systems (\ref{i})--(\ref{ix}) have
passed the Weiss-Kruskal algorithm well.

\section{Integrability\label{integrability}}

No two of the systems (\ref{i})--(\ref{ix}) can be related by transformations
(\ref{3}), but not all of (\ref{i})--(\ref{ix}) are distinct with respect to
more general transformations. If we take (\ref{i}) with $k=0$ and make
$v_{x}=w$, then we get exactly (\ref{viii}) for $u$ and $w$. If we make
$v_{x}=w$ in (\ref{ii}), then we get exactly (\ref{vii}) for $u$ and $w$. If
we make $v_{x}=w$ in (\ref{vi}), then we get (\ref{ix}) with $k=0$ for~$u$ and
$w$. Therefore we can restrict our further study to (\ref{i})--(\ref{v}) and
(\ref{ix}) only.

Since the systems (\ref{i})--(\ref{v}) and (\ref{ix}) pass the Painlev\'{e}
test well, they can be expected to be integrable. Let us try to f\/ind their Lax
pairs. We consider the over-determined linear system
\begin{equation}
\Psi_{x}=A\Psi, \qquad \Psi_{t}=B\Psi, \label{9}
\end{equation}
where $A$ and $B$ are some matrices and $\Psi$ is a column, and require that
the compatibility condition of (\ref{9}),
\begin{equation}
A_{t}=B_{x}-AB+BA, \label{10}
\end{equation}
represents the system under consideration. We assume that
\begin{equation}
A=Pu+Qv+R \label{11}
\end{equation}
and $B=B(u$, $u_{x}$, $u_{xx}$, $v$, $v_{x}$, $v_{xx})$, where $P$, $Q$ and
$R$ are constant matrices. Under this assumption, we get from (\ref{10}) an
explicit expression for $B$ in terms of $P$, $Q$, $R$ and a constant matrix
$S$, as well as a set of conditions for $P$, $Q$, $R$ and $S$. Then we try to
satisfy those conditions, increasing the dimension of the matrices. (This is
essentially the Dodd-Fordy version \cite{DF} of the Wahlquist-Estabrook method
\cite{WE}.) In this way, we obtain the following results for the systems
(\ref{i}), (\ref{ii}), (\ref{iii}), (\ref{iv}) and (\ref{ix}):
\begin{equation}
A_{(\mbox{\scriptsize i})}=\left(
\begin{array}{cccc}
0 & u+\sigma & 0 & kv\\
2 & 0 & 0 & 0\\
0 & v & 0 & u-\sigma\\
0 & 0 & 2 & 0
\end{array}
\right) , \label{12}
\end{equation}
$B_{(\mbox{\scriptsize i})}=\{\{-2u_{x}$, $u_{xx}-4u^{2}+8kv^{2}+4\sigma u+8\sigma^{2}$,
$4kv_{x}$, $-2kv_{xx}+4kuv\}$, $\{-8u+16\sigma$, $2u_{x}$, $16kv$,
$-4kv_{x}\}$, $\{4v_{x}$, $-2v_{xx}+4uv$, $-2u_{x}$, $u_{xx}-4u^{2}
+8kv^{2}-4\sigma u+8\sigma^{2}\}$, $\{16v$, $-4v_{x}$, $-8u-16\sigma$,
$2u_{x}\}\}$;
\begin{equation}
A_{(\mbox{\scriptsize ii})}=\left(
\begin{array}{ccc}
2\sigma &  u & 0\\
2 & -\sigma & 0\\
v & 0 & -\sigma
\end{array}
\right), \label{13}
\end{equation}
$B_{(\mbox{\scriptsize ii})}=\{\{-2u_{x}-6\sigma u+18\sigma^{3}$, $u_{xx}-4u^{2}+3\sigma
u_{x}+9\sigma^{2}u$, $0\}$, $\{-8u+18\sigma^{2}$, $2u_{x}+6\sigma
u-9\sigma^{3}$, $0\}$, $\{v_{xx}-4uv-3\sigma v_{x}+9\sigma^{2}v$,
$vu_{x}-uv_{x}+3\sigma uv$, $-9\sigma^{3}\}\}$;
\begin{equation}
A_{(\mbox{\scriptsize iii})}=\left(
\begin{array}{ccc}
0 & u+\sigma & 0\\
2 & 0 & 0\\
0 & v & 0
\end{array}
\right), \label{14}
\end{equation}
$B_{(\mbox{\scriptsize iii})}=\{\{-2u_{x}$, $u_{xx}-4u^{2}+4\sigma u+8\sigma^{2}$, $0\}$,
$\{-8u+16\sigma$, $2u_{x}$, $0\}$, $\{-8v_{x}$, $4v_{xx}-4uv+8\sigma v$,
$0\}\}$;
\begin{equation}
A_{(\mbox{\scriptsize iv})}=\left(
\begin{array}{ccc}
\sigma &  u-v & \sigma(\frac{8}{3}u-4v)\\
\frac{3}{2} & -2\sigma &  u-v\\
0 & \frac{3}{2} & \sigma
\end{array}
\right), \label{15}
\end{equation}
$B_{(\mbox{\scriptsize iv})}=\{\{6v_{x}-3\sigma(u-3v)-18\sigma^{3}$, $-4v_{xx}
+6uv-6v^{2}+2\sigma(u_{x}-3v_{x})+6\sigma^{2}(u-3v)$, $\sigma(-\frac{4}
{3}u_{xx}-4v_{xx}+6u^{2}+4uv-18v^{2})\}$, $\{9v-27\sigma^{2}$, $6\sigma
(u-3v)+36\sigma^{3}$, $-4v_{xx}+6uv-6v^{2}-2\sigma(u_{x}-3v_{x})+6\sigma
^{2}(u-3v)\}$, $\{\frac{27}{2}\sigma$, $9v-27\sigma^{2}$, $-6v_{x}
-3\sigma(u-3v)-18\sigma^{3}\}\}$;
\begin{equation}
A_{( \mbox{\scriptsize ix})}=\left(
\begin{array}{cccc}
0 & u+\sigma & 0 & 0\\
2 & 0 & 0 & 0\\
0 & v+\tau & 0 & u+\sigma\\
-2k & 0 & 2 & 0
\end{array}
\right), \label{16}
\end{equation}
$B_{(\mbox{\scriptsize ix})}=\{\{-2u_{x}$, $u_{xx}-4u^{2}+4\sigma u+8\sigma^{2}$, $0$,
$0\}$, $\{-8u+16\sigma$, $2u_{x}$, $0$, $0\}$, $\{-2v_{x}$, $ku_{xx}
+v_{xx}-8uv+4\tau u+4\sigma v+16\sigma\tau$, $-2u_{x}$, $u_{xx}-4u^{2}+4\sigma
u+8\sigma^{2}\}$, $\{8ku-8v-16k\sigma+16\tau$, $2v_{x}$, $-8u+16\sigma$,
$2u_{x}\}\}$; where the index of $A$ and $B$ corresponds to the system,
$\sigma$ and $\tau$ are arbitrary parameters, and the cumbersome matrices $B$
are written by rows. The invariant technique from \cite{S95} allows us to
prove that the parameters $\sigma$ and $\tau$ are essential, i.e. they cannot
be eliminated by gauge transformations of $\Psi$.

The system (\ref{i}) is the original Hirota-Satsuma equation \cite{HS}, and
the Lax pair found in~\cite{DF} corresponds to our result (\ref{9}) and
(\ref{12}) with $k=1$ up to a necessary transformation (\ref{3}) and a gauge
transformation of $\Psi$. The system (\ref{ix}) represents the two
$(1+1)$-dimensional reductions \cite{Sak} of the $(2+1)$-dimensional perturbed KdV
equation \cite{MF}, and the Lax pair~(\ref{9}) and (\ref{16}) follows from the
$(2+1)$-dimensional Lax pair \cite{Sak} by reduction as well.

In the case of (\ref{v}), however, this approach leads us only to matrices $A$
containing no essential parameters, at least for dimensions from $2 \times$2 to
$5 \times$5; probably, the assumption~(\ref{11}) is too restrictive. The method
of truncation of singular expansions \cite{Wei} gives us a similar result:
though the truncation procedure turns out to be compatible for the system
(\ref{v}), the truncated singular expansions contain no essential parameters.
Therefore it remains unknown whether the system (\ref{v}) is integrable or not.

\section{Conclusion\label{conclusion}}

Let us make a brief comparison between our study of coupled KdV equations and
the classif\/ication given in \cite{Kar}.

\vspace{-2mm} 

\begin{itemize}
\item  The integrability of selected systems, equivalences between them, as
well as non-generic branches were not studied in \cite{Kar}.

\vspace{-2mm} 

\item  Karasu considered only the case $\alpha=\beta=-2$ (in our notations of
(\ref{4})) but did not use the condition $u_{0}v_{0}\neq0$; therefore the
results of \cite{Kar} could, in principle, contain all our systems with
$\beta\geq-2$, i.e. (\ref{i})-(\ref{viii}), and should miss only (\ref{ix})
with $\beta=-3$.

\vspace{-2mm} 

\item  Our systems (\ref{iii}), (\ref{v}) and (\ref{ix}) are missed in the
classif\/ication \cite{Kar}.

\vspace{-2mm} 

\item  Appropriate transformations (\ref{3}) change Karasu's systems
(13), (14), (15), (16), (17) and
(19) (numbered as in \cite{Kar}) into our systems (\ref{vi}),
(\ref{vii}), (\ref{viii}), (\ref{ii}), (\ref{i}) and (\ref{iv}), respectively,
eliminating all free parameters.

\vspace{-2mm} 

\item  The system (20) from \cite{Kar} can be transformed by
(\ref{3}) into a system of two non-coupled equations and therefore is absent
from our results.

\vspace{-2mm} 

\item  The system (21) from \cite{Kar} does not pass the Painlev\'{e}
test (even after a correction of misprints).
\vspace{-2mm} 

\end{itemize}

Below we give some unsolved problems.

\vspace{-2mm} 

\begin{itemize}
\item  Does the system (\ref{v}) possess a Lax pair with an essential parameter?

\vspace{-2mm} 

\item  Is there a system (\ref{1}) with the Painlev\'{e} property and
$\beta<-3$ in its singular generic branch?

\vspace{-2mm} 

\item  Can some of the systems (\ref{i})--(\ref{v}) and (\ref{ix}) be related
to each other by Miura and B\"{a}cklund transformations?
\vspace{-2mm} 

\end{itemize}

\label{sakovich-lp}
\end{document}